\documentclass[9pt,conference]{IEEEtran}
\usepackage{amssymb,amsthm,amsmath,array}
\usepackage{graphicx}
\usepackage[caption=false,font=footnotesize]{subfig}
\usepackage{xspace}
\usepackage[sort&compress, numbers]{natbib}
\usepackage{stmaryrd}
\usepackage{xcolor}
\usepackage{mathtools}
\usepackage{float}
\usepackage{textcomp}

\usepackage{amsmath,amssymb,amsthm}  

\usepackage{xcolor}

\definecolor{darkred}{rgb}{0.6,0,0}
\definecolor{darkgreen}{rgb}{0,0.5,0}
\definecolor{darkblue}{rgb}{0,0,0.5}
\usepackage{booktabs}
\usepackage{caption}
\usepackage[skins]{tcolorbox}
\usepackage{tikz,pgfplots}

\pgfplotsset{compat=1.9}
\usepgfplotslibrary{groupplots}
\usetikzlibrary{intersections,calc,arrows,matrix,spy,math,calc,fit,positioning}

\usepgfplotslibrary{statistics}
\usetikzlibrary{patterns}
\makeatletter
\tikzset{nomorepostaction/.code=\let\tikz@postactions\pgfutil@empty}
\makeatother

 \pgfplotsset{
        table/search path={figs/,figs/smlm_noise},
    }

\usetikzlibrary{arrows.meta,shapes.geometric}

\usepackage{multirow}

\usepackage[squaren,Gray]{SIunits}

\graphicspath{{figs/}}

\newcommand{\Nmeas}{M}

\newcommand{\uin}{u_{\mathrm{in}}}

\newcommand{\ii}{\mathrm{j}}

\newcommand{\bd}{\mathbf b}

\newcommand{\fd}{\mathbf f}

\newcommand{\pd}{\mathbf p}

\newcommand{\xd}{\mathbf x}
\newcommand{\yd}{\mathbf y}
\newcommand{\zd}{\mathbf z}

\newcommand{\Hd}{\mathbf H}

\usepackage{graphicx}
\graphicspath{{figs/}}
\usepackage[ruled]{algorithm}
\usepackage{algpseudocode,algorithmicx}



\def\vdotfill#1{\vtop to0pt{\null \dimen0=#1\baselineskip\advance\dimen0 by-.4ex 
   \kern-1.6ex \cleaders\hbox{\lower.4ex\vbox to1ex{}.}\vskip\dimen0 \vss}}
   
\algdef{SE}[FOR]{NoDoFor}{EndFor}[1]{\algorithmicfor\ #1}{\algorithmicend\ \algorithmicfor}%
\algdef{SE}[IF]{NoThenIf}{EndIf}[1]{\algorithmicif\ #1}{\algorithmicend\ \algorithmicif}%
\algdef{SE}[WHILE]{NoDoWhile}{EndWhile}[1]{\algorithmicwhile\ #1}{\algorithmicend\ \algorithmicwhile}%

\newcommand{\R}{{\mathbb R}}

\newcommand{\C}{\mathbb C}

\usepackage{mathtools}
\usepackage{xcolor} 

\begin{document}
\title{Optical Diffraction Tomography Meets Fluorescence Localization Microscopy}
\author{\IEEEauthorblockN{
        Thanh-An Pham\IEEEauthorrefmark{1}, 
        Emmanuel Soubies\IEEEauthorrefmark{2},  
        Ferréol Soulez\IEEEauthorrefmark{3}, and
        Michael Unser\IEEEauthorrefmark{4}
    }
    \IEEEauthorblockA{
        \IEEEauthorrefmark{1}3D Optical
Systems Group, Department of Mechanical Engineering, Massachusetts Institute of Technology, Cambridge,
USA.\\
        \IEEEauthorrefmark{2} IRIT, Université de Toulouse, CNRS, Toulouse, France.\\
        \IEEEauthorrefmark{3} Univ. Lyon, Univ. Lyon1, ENS de Lyon, CNRS, Centre de Recherche Astrophysique de Lyon, Saint-Genis-Laval, France.\\
        \IEEEauthorrefmark{4} Biomedical
Imaging Group, École polytechnique fédérale de Lausanne, Lausanne, Switzerland.
        }      
}
\maketitle
\begin{abstract}
We show that structural information can be extracted from single molecule localization microscopy (SMLM) data. More precisely, we reinterpret  SMLM data as the measures of a phaseless optical diffraction tomography system for which the illumination sources are fluorophores within the sample. Building upon this model, we  propose a joint optimization framework to estimate both the refractive index  map and the position of fluorescent molecules from the sole SMLM frames. 
\end{abstract}

\section{Introduction}

In fluorescence microscopy the heterogeneity of biological tissues---through variations in their refractive index (RI)---induces aberrations and light scattering, which distorts the recorded images. 
Beyond specific sample preparation such as tissue clearing, hardware and computational solutions have been proposed over the last two decades.
Adaptive optics and wavefront shaping~\cite{Gigan2017,ahn2019overcoming,Yoon2020} emerged as powerful hardware strategies allowing to correct the distorted wavefront. 
Alternatively, a few multimodal systems---capable of acquiring both fluorescence and phase data---have been developed~\cite{Chung2016,Shin2018}. Not only do they allow to reconstruct the RI map (from phase measurements) and to exploit it to improve the fluorescence signal, but they also offer a unique combination of \textit{structural} (RI) and \textit{functional} (fluorescence) information about the sample. 

In an attempt to simplify the acquisition process, two recent works~\cite{Xue22,Pham2021b} addressed the problem of reconstructing both  RI and fluorophore density from the same data collected by standard fluorescence systems (i.e., without phase measurements). Both works exploit the fact that fluorescence images can be seen as measurements of a phaseless optical-diffraction tomography (ODT) system~\cite{tian20153d,Pham2018}, where the illumination sources are the emission of fluorophores inside the sample. Indeed, fluorophores emit light that scatters through the sample before being captured by the camera, which shows that the recorded fluorescence images carry information on the RI of the sample. This structural information can then be unveiled using dedicated numerical methods~\cite{Xue22,Pham2021b}.
In~\cite{Xue22}, the authors exploited scanning microscopy to localize the activation of fluorophores, while in~\cite{Pham2021b}, we capitalized on the single molecule activation property of~SMLM.

  \paragraph*{Outline} 
  The purpose of this communication is to present and explain the main steps of a computational pipeline that simultaneously localizes fluorophores embedded in a specimen while reconstructing its refractive index.
  In Section~\ref{sec:fwd}, we show how an SMLM image can be interpreted as phaseless ODT measurements. We end up with an image formation model that depends on both the RI map and the positions and amplitudes of fluorescent molecules. Then, we present in Section~\ref{sec:Optim} our joint optimization framework. Finally, we present some numerical results in Section~\ref{sec:num}.

\section{The ODT from SMLM Forward Model}\label{sec:fwd}

Let $\Omega \subseteq \R^3$ be the image domain and $\eta:\Omega\rightarrow \R$ the RI distribution of the sample. Moreover, let $\uin(\cdot;  \pd, a) : \Omega \to \C$ denote the spherical wave emitted by a fluorescent molecule with  intensity~$a>0$ located at position $\pd \in \Omega$. More precisely, we have
\begin{equation}
\label{eq:spherical}
  \forall \xd \in \Omega, \;   \uin(\xd; \pd,a) = a\frac{\exp{\left(\ii k_\mathrm{b} \| \xd -  \pd\|_2\right)}}{4\pi \| \xd -  \pd\|_2},
\end{equation}
where~$k_\mathrm{b}={2\pi \eta_\mathrm{b}}/{\lambda}$ is the wavenumber with $\lambda$ the emission wavelength and $\eta_\mathrm{b} > 1$ the RI  of the surrounding medium. 

 The resulting SMLM frame $\yd \in \R^\Nmeas$ can then be described by
\begin{equation}\label{eq:model_cont}
    y_m =  \mathrm{Pois}\left(\left|\left(P \,u_{\mathrm{t}}\left(\cdot;  \pd, a\right)\right)(\xd_m)\right|^2 + b_m \right),
\end{equation}
where $\mathrm{Pois}$ denotes Poisson's distribution (shot noise), $\{\xd_m\}_{m=1}^\Nmeas$ are the camera sampling points,   $\bd\in\R^\Nmeas$ is a background fluorescent signal, and $P $ is a linear integral operator that models the effect of the optical system. Finally, 
$u_{\mathrm{t}}\left(\cdot;  \pd, a\right) :\Omega \to \C$ represents the total field, resulting from the scattering of the spherical wave through the sample. It is governed by the  Lippmann-Schwinger equation
\begin{multline}
\label{eq:lipp}
    u_{\mathrm{t}}\left(\xd;  \pd, a\right) = \uin(\xd;\pd,a)  + \int_\Omega g(\xd - \zd) f(\zd) u_{\mathrm{t}}\left(\zd;  \pd, a\right) \, \mathrm{d}\zd,
\end{multline}
where $f(\xd) = k_\mathrm{b}^2\left({\eta(\xd)^2}/{\eta_\mathrm{b}^2} - 1\right)$ is the scattering potential and
$g:\R^3\rightarrow\C$ the Green function given by  $g = \uin(\cdot;\mathbf{0},1)$~\citep{cornea2014fluorescence}.

  Note that, because fluorophores are incoherent sources, the image~$\yd$ formed out of the activation of $L$ molecules at positions $\{\pd_l\}_{l=1}^L$ with amplitudes $\{a_l\}_{l=1}^L$ is simply given by $\yd=\sum_{l=1}^L \yd_l$, where $\yd_l$ stands for the contribution of the $l$th fluorophore. As such, without loss of generality, we assume  to  have access to  $L$ SMLM acquisitions $\{\yd_l\}_{l=1}^L$, each corresponding to the activation of one molecule.

  The discretization of this forward model (i.e.,  equations~\eqref{eq:model_cont} and~\eqref{eq:lipp}), requires to deal with the singularity at 0 of the integrand in~\eqref{eq:lipp}. We refer the reader to \cite{pham2020three} for details on how this difficulty can be efficiently handled. In the following, we denote the (nonlinear) discrete forward model by $\Hd(\fd, \pd,a)$, where $\fd \in \R^N$ corresponds to a sampled version of the scattering potential~$f$ within~$\Omega$.
  
\section{RI Reconstruction and Molecule Localization}\label{sec:Optim}

  Given the SMLM data $\{\yd_l\}_{l=1}^L$, we propose to estimate both the RI map $\fd$ and the molecules positions $\{\pd_l\}_{l=1}^L$ and amplitudes $\{a_l\}_{l=1}^L$ through the resolution of
  \begin{equation}\label{eq:Pb_Opti}
  \underset{\fd \geq \mathbf{0}, \,  \pd_l \in \Omega, \, a_l >0}{\mathrm{arg \; min}} \; \sum_{l=1}^L \mathcal{D}_{\mathrm{KL}}\left(\Hd(\fd,\pd_l,a_l) + \bd_l;\yd_l \right)  + \tau\|\nabla \fd\|_{2,1},
\end{equation}
where $\tau >0$, $\|\nabla \fd\|_{2,1}$ is the popular total-variation~(TV) regularization, and $\mathcal{D}_{\mathrm{KL}}$ stands for the generalized Kullback-Leibler divergence. It is  defined  by
$
    \mathcal{D}_{\mathrm{KL}}\left(\zd;\yd\right) = \zd^T \mathbf{1}_M - \yd \odot \log(\zd + \beta),
$
with $\beta > 0$. From a Bayesian point of view, it corresponds (up to a constant term) to the negative log-likelihood associated with the Poisson distribution (shot noise). Finally, $\bd_l$ models the background fluorescence of the $l$th SMLM frame $\yd_l$. In practice, we estimate it in a pre-processing step, exploiting the fact  that it varies slowly in space and time~\cite[Section 3.4.3]{Pham2021b}. \\

To tackle Problem~\eqref{eq:Pb_Opti}, we deploy an alternating optimization strategy. More precisely, we sequentially solve the following subproblems.
\begin{itemize}
    \item \textit{Update of molecules amplitudes $\{a_l\}_{l=1}^L$.} Using the fact that $\Hd(\fd,\pd_l,a_l) = a_l^2 \Hd(\fd,\pd_l,1)$, we  obtain simple expressions of the first two derivatives of the objective function in~\eqref{eq:Pb_Opti} with respect to the amplitudes $a_l$. This allows us to deploy an efficient Newton update.
    \item \textit{Update of molecules positions $\{\pd_l\}_{l=1}^L$.} The non-differentiability of the spherical wave in~\eqref{eq:spherical}  at $\xd = \pd$ prevents the direct use of a gradient-based approach on~\eqref{eq:Pb_Opti}. Instead, we consider a smoothed version of the spherical wave where the norms in~\eqref{eq:spherical} are replaced by $\|\cdot\|_{2,\epsilon}= \sqrt{\|\cdot\|_2^2 + \epsilon}$ with $0<\epsilon \ll 1$.  We then derive a closed-form expression of the gradient of the objective with respect to each $\pd_l$, and set up a projected gradient method (to ensure that $\pd_l \in \Omega$). 
    \item \textit{Update of refractive index $\fd$.} This corresponds to a phaseless inverse-scattering problem~\cite{tian20153d,Pham2018}. We address it through a relaxed variant of FISTA~\cite{ma2018accelerated}.
\end{itemize}
Due to space limitations, we refer the reader to~\cite[Section 3]{Pham2021b} for a detailed description of theses three steps. Finally, as the objective in~\eqref{eq:Pb_Opti} is non-convex, initialization plays a central role. For molecules amplitudes and positions, initial estimates can be obtained with any SMLM algorithm. Concerning the initialization of the refractive index $\fd$, we simply consider a rescaled version of the widefield image.

\section{Numerical Illustration}\label{sec:num}

\paragraph*{Simulation Setting}
We simulated an RI distribution immersed in water and included in the region $\Omega$ of size~$(7.2\times7.2\times3.2) \micro\meter^3$. Fluorophores have then been randomly distributed within this sample. The smallest distance between two fluorophores is~$20\nano\meter$, the amplitudes~$a_l$ were drawn from a Poisson distribution with mean~$A=1000$, and the emission wavelength was set to $\lambda = 647 \nano\meter$.
We then generated a number of $L=1000$ SMLM frames (biplane modality), each corresponding to the activation of a single fluorophore. Finally, the backgrounds $\bd_l$ were simulated through noise smoothing (for slow space variation) and interpolation (for slow time variation).

\paragraph*{Results and Discussion}
In Fig.~\ref{fig} (top), we compare the RI map obtained by the proposed joint optimization framework with the ones that  obtained when the molecules positions and amplitudes are fixed to their i) initial estimates or ii) true values (best-case scenario). These results highlight the importance of the joint optimization, as well as its efficiency.
In the bottom part of Fig.~\ref{fig}, we compare  molecules amplitudes and positions estimated by our joint optimization approach with initial ones. We can  appreciate the gain in accuracy (twice as good RMSE on localization). Through the estimation of the RI variations, we account for sample-induced distortions and improve the accuracy of the localization of molecules.

  More extensive experiments are reported in~\cite{Pham2021b}. In particular, we show that the  proposed approach is robust to noise and still performs  well when reducing the number of frames to $L=100$. We also illustrate   how the distribution of fluorophores affects the quality of the reconstructed RI map.

\newcommand{\winszxy}{72}
\newcommand{\winszz}{32}
\newcommand{\insep}{0.5pt}
\newcommand{\widthr}{0.12}
\newcommand{\foldoi}{./}

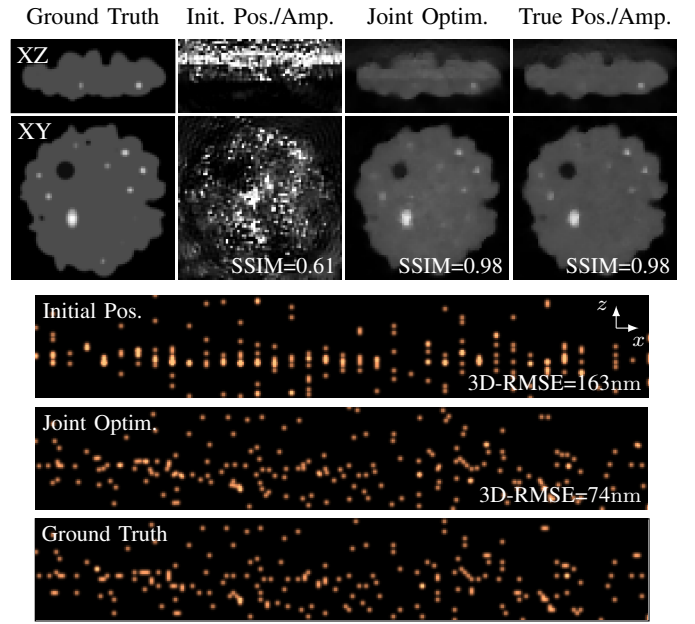
\begin{figure}
    \centering
    \begin{tikzpicture}
    \begin{axis}[enlargelimits=false, axis equal image, width=\widthr\textwidth,
    scale only axis, 
    xmin=0,xmax=72,ymin=0,ymax=32,
    at={(0,0)},anchor=west,title={Ground Truth}, 
    every axis title/.style={yshift={4*\insep},anchor=south,at={(axis cs:36,33)}},
    ]
    \pgfplotsset{ticks=none}
    \addplot graphics [xmin=0,xmax=72,ymin=0,ymax=32,includegraphics={width=\widthr\textwidth}] {{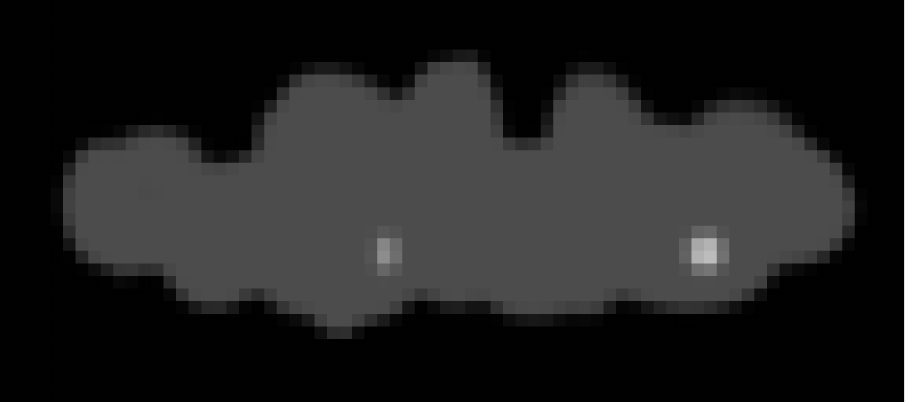}};
    \node[inner sep=\insep] (XZgt) at (axis cs:72,16) {};
    \node[inner sep=\insep] (XZgtb) at (axis cs:36,0) {};
     \node[anchor=south west,white] at (axis cs:-1.3,18) {\color{white}XZ};
    \end{axis}
    
    \node[inner sep=\insep,anchor = north] (XY2GT) at (XZgtb.south)
    {\includegraphics[width=\widthr\textwidth]{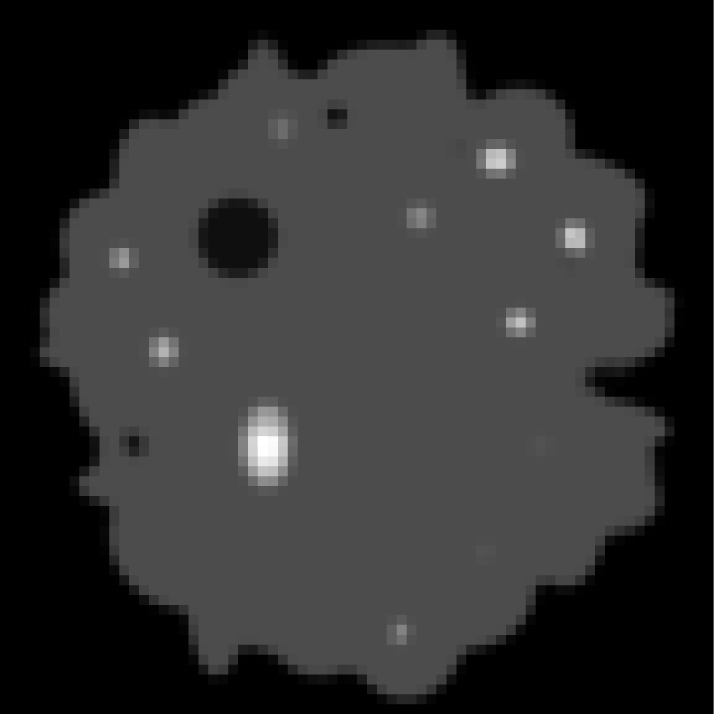}};
    \node[anchor=north west,color=white] at (XY2GT.north west) {XY};

    \node[inner sep=\insep,anchor=west] (XZbad) at (XZgt.east)
    {\includegraphics[width=\widthr\textwidth]{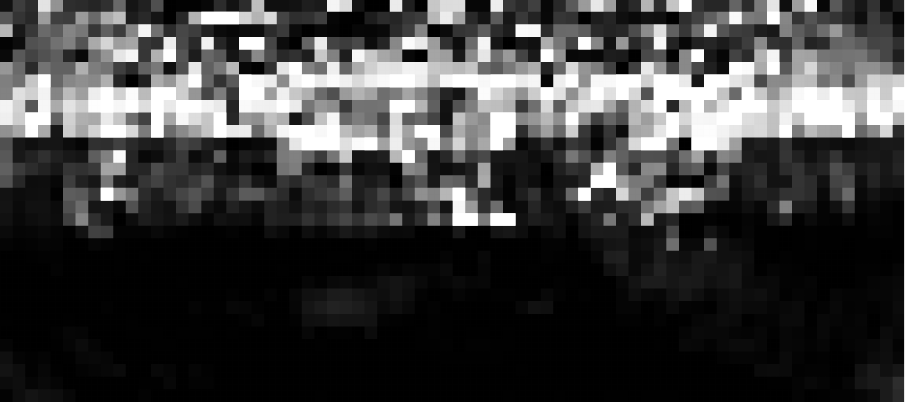}};\node[anchor=south] at (XZbad.north) {Init. Pos./Amp.};

    \node[inner sep=\insep,anchor=west] (XY2bad) at (XY2GT.east)
    {\includegraphics[width=\widthr\textwidth]{ 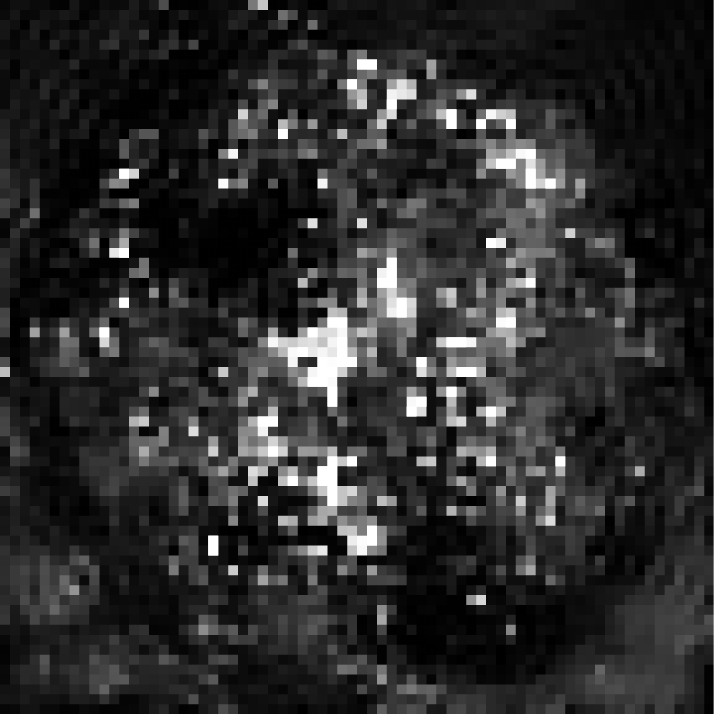}};
    \node[anchor=south east,color=white] at (XY2bad.south east) {\small SSIM=0.61};

    \node[inner sep=\insep,anchor=west] (XY2joint) at (XY2bad.east)
    {\includegraphics[width=\widthr\textwidth]{ 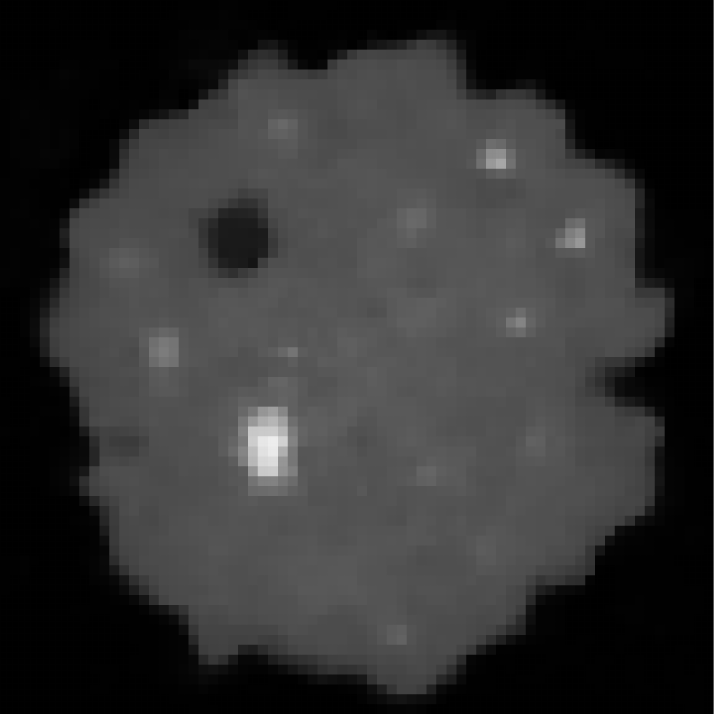}};
    \node[anchor=south east,color=white] at (XY2joint.south east) {\small SSIM=0.98};
    
\node[inner sep=\insep,anchor=west] (XZjoint) at (XZbad.east)
    {\includegraphics[width=\widthr\textwidth]{ 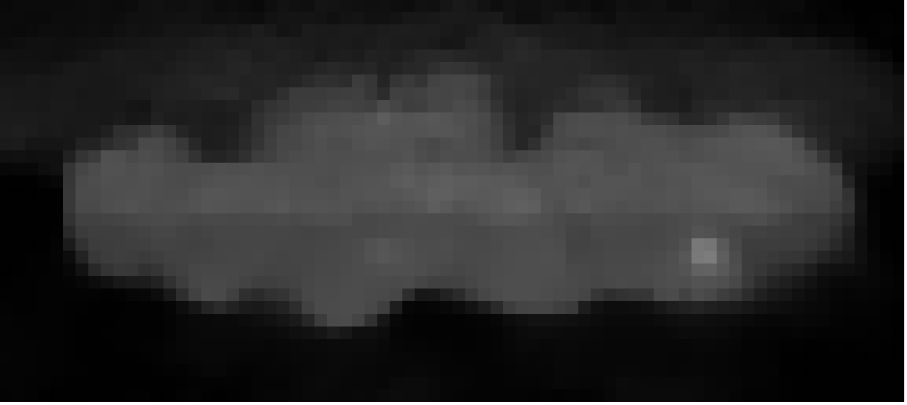}};\node[anchor=south] at (XZjoint.north) {Joint Optim.};

    \node[inner sep=\insep,anchor=west] (XY2best) at (XY2joint.east)
    {\includegraphics[width=\widthr\textwidth]{ 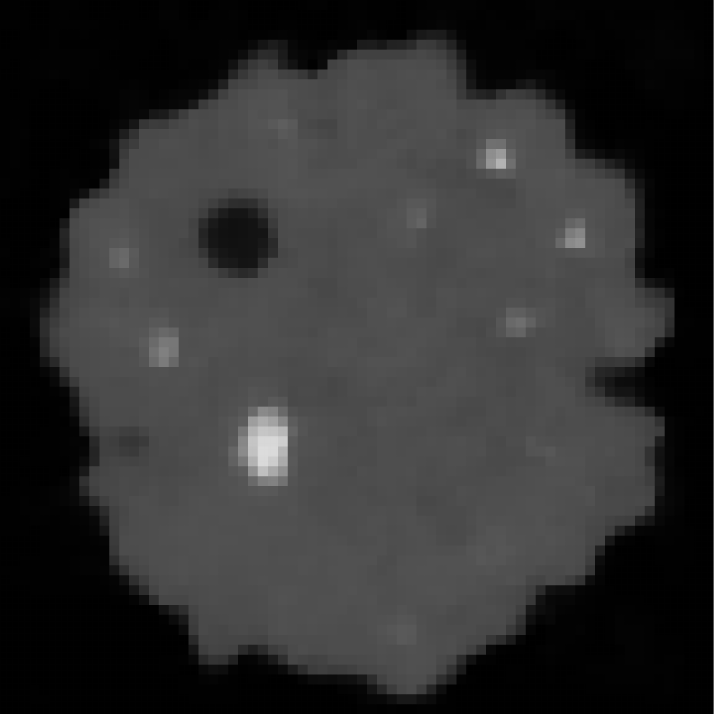}};
    \node[anchor=south east,color=white, align=right] at (XY2best.south east) {\small SSIM=0.98};

\node[inner sep=\insep,anchor=west] (XZbest) at (XZjoint.east)
    {\includegraphics[width=\widthr\textwidth]{ 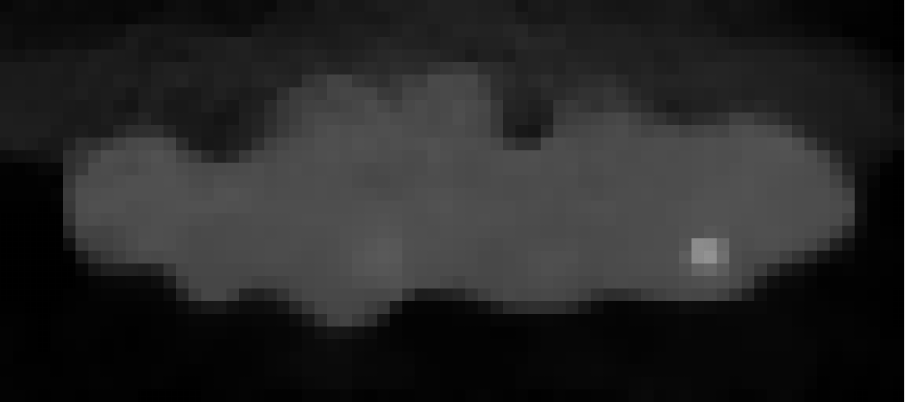}};\node[anchor=south] at (XZbest.north) {True {Pos./Amp.}};

\renewcommand{\insep}{3}
\newcommand{\intersep}{0}
\renewcommand{\widthr}{0.45}
\newcommand{\shiftCSx}{-0.55}
\newcommand{\shiftCSy}{-0.55}

     \node[anchor=north,inner sep=\insep] (bad) at (4.4,-2.8)
    {\includegraphics[width=\widthr\textwidth]{ 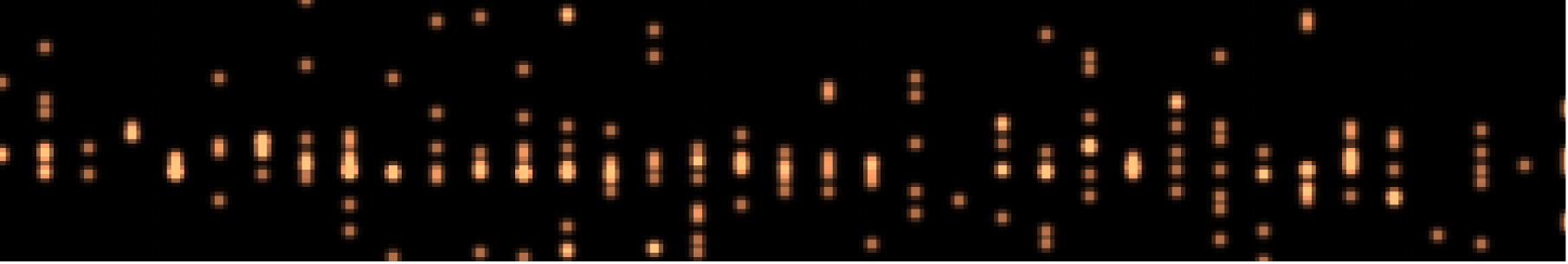}}; 
    \node[anchor=north west,white] at ($(bad.north west) + (0.1,-0.1)$) {\small Initial Pos.};
    \node[anchor=south east,color=white] at ($(bad.south east)+(-0.1,0.1)$) {\small 3D-RMSE=163\nano\meter};
    
    \node[anchor=north,inner sep=\insep] (joint) at ($(bad.south) + (0,3pt)$)
    {\includegraphics[width=\widthr\textwidth]{ 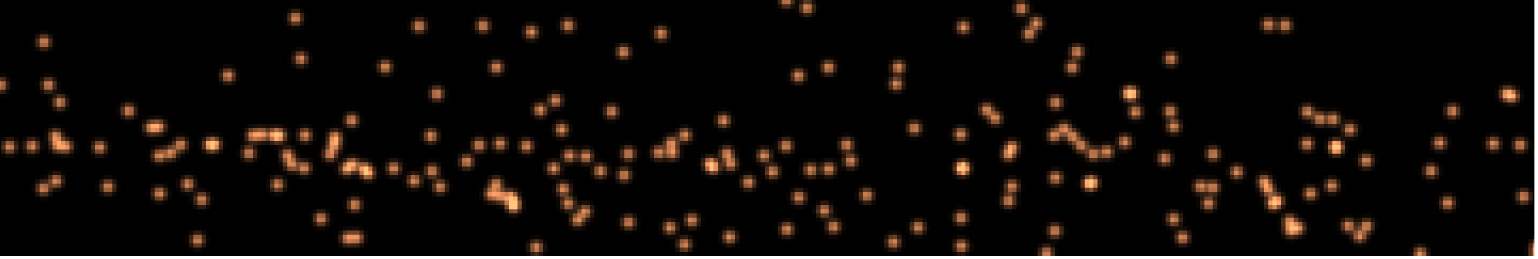}};
    \node[anchor=north west,white] at ($(joint.north west) + (0.1,-0.1)$) {\small Joint Optim.};
    \node[anchor=south east,color=white] at ($(joint.south east)+(-0.1,0.1)$) {\small 3D-RMSE=74\nano\meter};

    \begin{axis}[enlargelimits=false, axis equal image, width=\widthr\textwidth,
    scale only axis ,at = {($(joint.south) - (0,\intersep)$)}, name={GT},inner sep=\insep,anchor=north,xmin=0,xmax=36,ymin=0,ymax=6]
    \pgfplotsset{ticks=none}
    \addplot graphics [xmin=0,xmax=36,ymin=0,ymax=6,includegraphics={width=\widthr\textwidth}] { 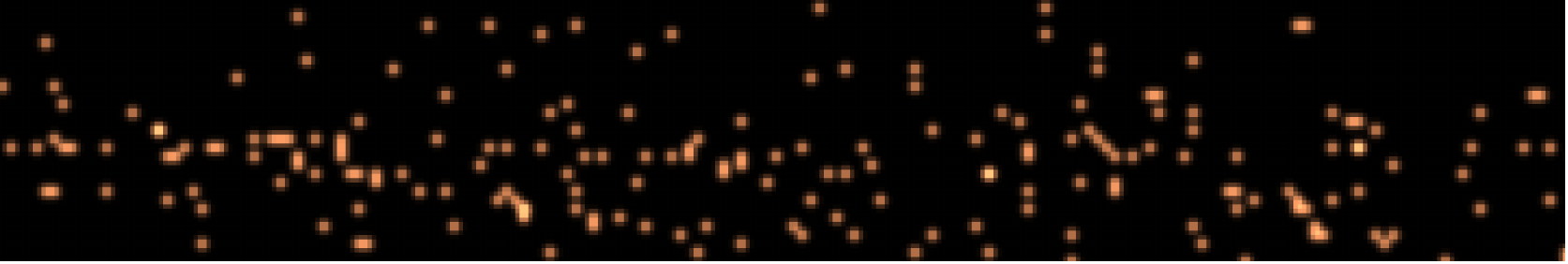};
    \node[white,anchor=north west] at (axis cs:-0.1,6.1) {\small Ground Truth};
    \end{axis}
    
\draw[-latex,white] ($(bad.north east) + (\shiftCSx,\shiftCSy)$) -- ($(bad.north east) + (\shiftCSx,\shiftCSy) + (0,0.3)$) node[anchor=east,white] {\small $z$};
    \draw[-latex,white] ($(bad.north east) + (\shiftCSx,\shiftCSy)$) -- ($(bad.north east) + (\shiftCSx,\shiftCSy) + (0.3,0)$) node[anchor=north,white] {\small $x$};
    
    \end{tikzpicture}
    \caption{\label{fig} Reconstructions of the RI volume (Top) and projection along Y of the rendered fluorescent molecules (Bottom).}
\end{figure}

\bibliographystyle{IEEEtran}
\bibliography{references}

\begin{thebibliography}{10}
\providecommand{\url}[1]{#1}
\csname url@samestyle\endcsname
\providecommand{\newblock}{\relax}
\providecommand{\bibinfo}[2]{#2}
\providecommand{\BIBentrySTDinterwordspacing}{\spaceskip=0pt\relax}
\providecommand{\BIBentryALTinterwordstretchfactor}{4}
\providecommand{\BIBentryALTinterwordspacing}{\spaceskip=\fontdimen2\font plus
\BIBentryALTinterwordstretchfactor\fontdimen3\font minus
  \fontdimen4\font\relax}
\providecommand{\BIBforeignlanguage}[2]{{%
\expandafter\ifx\csname l@#1\endcsname\relax
\typeout{** WARNING: IEEEtran.bst: No hyphenation pattern has been}%
\typeout{** loaded for the language `#1'. Using the pattern for}%
\typeout{** the default language instead.}%
\else
\language=\csname l@#1\endcsname
\fi
#2}}
\providecommand{\BIBdecl}{\relax}
\BIBdecl

\bibitem{Gigan2017}
S.~Gigan, ``Optical microscopy aims deep,'' \emph{Nature Photonics}, vol.~11,
  2017.

\bibitem{ahn2019overcoming}
C.~Ahn, B.~Hwang, K.~Nam, H.~Jin, T.~Woo, and J.-H. Park, ``Overcoming the
  penetration depth limit in optical microscopy: Adaptive optics and wavefront
  shaping,'' \emph{Journal of Innovative Optical Health Sciences}, vol.~12,
  2019.

\bibitem{Yoon2020}
S.~Yoon, M.~Kim, M.~Jang, Y.~Choi, W.~Choi, S.~Kang, and W.~Choi, ``Deep
  optical imaging within complex scattering media,'' \emph{Nature Reviews
  Physics}, vol.~2, 2020.

\bibitem{Chung2016}
J.~Chung, J.~Kim, X.~Ou, R.~Horstmeyer, and C.~Yang, ``Wide field-of-view
  fluorescence image deconvolution with aberration-estimation from fourier
  ptychography,'' \emph{Biomedical optics express}, vol.~7, 2016.

\bibitem{Shin2018}
S.~Shin, D.~Kim, K.~Kim, and Y.~Park, ``Super-resolution three-dimensional
  fluorescence and optical diffraction tomography of live cells using
  structured illumination generated by a digital micromirror device,''
  \emph{Scientific reports}, vol.~8, 2018.

\bibitem{Xue22}
Y.~Xue, D.~Ren, and L.~Waller, ``Three-dimensional bi-functional refractive
  index and fluorescence microscopy (brief),'' \emph{Biomed. Opt. Express},
  vol.~13, 2022.

\bibitem{Pham2021b}
T.~an~Pham, E.~Soubies, F.~Soulez, and M.~Unser, ``Optical diffraction
  tomography from single-molecule localization microscopy,'' \emph{Optics
  Communications}, vol. 499, 2021.

\bibitem{tian20153d}
L.~Tian and L.~Waller, ``3{D} intensity and phase imaging from light field
  measurements in an {LED} array microscope,'' \emph{Optica}, vol.~2, 2015.

\bibitem{Pham2018}
T.-A. Pham, E.~Soubies, A.~Goy, J.~Lim, F.~Soulez, D.~Psaltis, and M.~Unser,
  ``Versatile reconstruction framework for diffraction tomography with
  intensity measurements and multiple scattering,'' \emph{Optics express},
  vol.~26, 2018.

\bibitem{cornea2014fluorescence}
A.~Cornea and P.~M. Conn, \emph{\textit{Fluorescence Microscopy:
  Super-Resolution and Other Novel Techniques}}.\hskip 1em plus 0.5em minus
  0.4em\relax Elsevier, 2014.

\bibitem{pham2020three}
T.-a. Pham, E.~Soubies, A.~Ayoub, J.~Lim, D.~Psaltis, and M.~Unser,
  ``{Three-dimensional optical diffraction tomography with Lippmann-Schwinger
  model},'' \emph{IEEE Transactions on Computational Imaging}, vol.~6, 2020.

\bibitem{ma2018accelerated}
Y.~Ma, H.~Mansour, D.~Liu, P.~T. Boufounos, and U.~S. Kamilov, ``Accelerated
  image reconstruction for nonlinear diffractive imaging,'' in \emph{IEEE
  ICASSP}, Calgary AB, Canada, 2018.

\end{thebibliography}
\end{document}